\documentclass[sigconf]{acmart}
\usepackage{booktabs}
\usepackage[ruled,vlined]{algorithm2e}
\usepackage{amsmath}

\DeclareMathAlphabet{\altmathcal}{OMS}{cmsy}{m}{n}
\usepackage{bm}
\usepackage{amsfonts}
\usepackage{enumitem}
\usepackage{tabularx}

\usepackage{float} 
\usepackage{multirow}
\usepackage{makecell}
\usepackage{arydshln}
\newcommand{\dataset}{\textsc{GUIDE}}
\newcommand{\methodshort}{\textsc{AIP}}
\newcommand{\method}{\textsc{Adaptive Incident Prioritization}}
\newcommand{\platform}{Microsoft Defender}

\newcommand{\hide}[1]{}

\usepackage{tikz}
\usetikzlibrary{arrows.meta, positioning, fit, shapes.geometric}

\AtBeginDocument{%
  }

\acmConference[arXiv '26]
{arXiv}
{2026}
{}

\begin{document}

\title{Adaptive Incident Prioritization for Security Operations at Scale}

\author{Scott Freitas}
\authornote{Both authors contributed equally to this work.}
\affiliation{%
  \institution{Microsoft Security Research}
  \state{Arizona}
  \country{USA}}
\email{scottfreitas@microsoft.com}

\author{Amir Gharib}
\authornotemark[1]
\affiliation{%
  \institution{Microsoft Security Research}
  \city{Toronto}
  \country{Canada}
}
\email{agharib@microsoft.com}

\author{Maayan Magenheim}
\affiliation{%
  \institution{Microsoft Security}
  \city{Herzliya}
  \country{Israel}
}
\email{mmagenheim@microsoft.com}

\renewcommand{\shortauthors}{Scott Freitas, Amir Gharib, \& Maayan Magenheim}

\begin{abstract}
Large security operations centers (SOCs) often face hundreds of active incidents per day, creating substantial cognitive and operational load for analysts. These analysts must quickly decide which incidents deserve attention from long, constantly changing queues, yet incidents are often ordered by arrival time, coarse severity, or product-specific heuristics, leaving priority unclear.
We introduce \method{} (\methodshort{}), the ranking algorithm behind \platform{} Queue Assistant, which continuously prioritizes security incidents for analyst investigation.
\methodshort{} adapts BM25-style ranking to a query-less, multi-tenant queue setting by representing each incident as a bag of normalized security components extracted from alerts and metadata. The model combines saturated local component frequency, global component rarity estimated over a cross-tenant corpus, bounded domain-prior multipliers, and component-level explanations. Deployed across tens of thousands of customers, \methodshort{} runs near-real-time inference that refreshes incident scores with 5-second median latency.
\methodshort{} achieves 92.8\% P@10 in expert-reviewed evaluation across 1,000 customer organizations. In post-launch telemetry across 473K organization-day queues, \methodshort{} increases alert-detail interaction by 5.8\% and alert-detail view events by 17.5\% relative to severity ordering, providing behavioral evidence that \methodshort{} concentrates analyst engagement.
We also extend the Microsoft \dataset{} dataset with the first public label source for SOC queue prioritization over real-world incidents, covering 499 organization queues and 9,980 incidents with expert priority labels. This resource enables the research community to develop, compare, and advance incident prioritization research. 
\end{abstract}




\maketitle

\section{Introduction}
Enterprise security has advanced substantially, yet a fundamental operational gap remains: SOCs receive more incidents than analysts can investigate. Prior work reports that large organizations can see up to 232 incidents per day, 51\% of which are never resolved~\cite{freitas2026genai}. Unified security operation platforms, such as \platform{}, consolidate enterprise alerts into cohesive incident queues across heterogeneous security products~\cite{einav2023introducing,freitas2024graphweaver}. However, consolidation alone does not answer the analyst's most urgent operational question: \emph{which incident should be investigated first?} 
Existing queue-ordering policies are insufficient for this prioritization task: arrival-time ordering rewards recency rather than impact, severity ordering is often too coarse and inconsistent across products, and heuristic priority rules are brittle as detector coverage, product integrations, and attacker behavior evolve.

This prioritization problem is distinct from incident triage: the goal is not to determine whether an incident is true or false positive, but to rank active incidents by investigation urgency. 
A true-positive incident may still be low priority if it is informational or low impact compared to competing incidents. These limitations motivate \methodshort{}, the ranking architecture behind Queue Assistant~\cite{gharib2026introducing,microsoft2026queueassistant}. \methodshort{} adapts BM25-style ranking to a query-less, multi-tenant SOC queue setting by representing incidents with interpretable security components that capture detection, threat, and potential impact. 
This keeps priority explainable while avoiding direct dependence on sparse or biased operational signals that do not directly measure queue-relative urgency.

\begin{figure*}
\centering
\includegraphics[width=\textwidth]{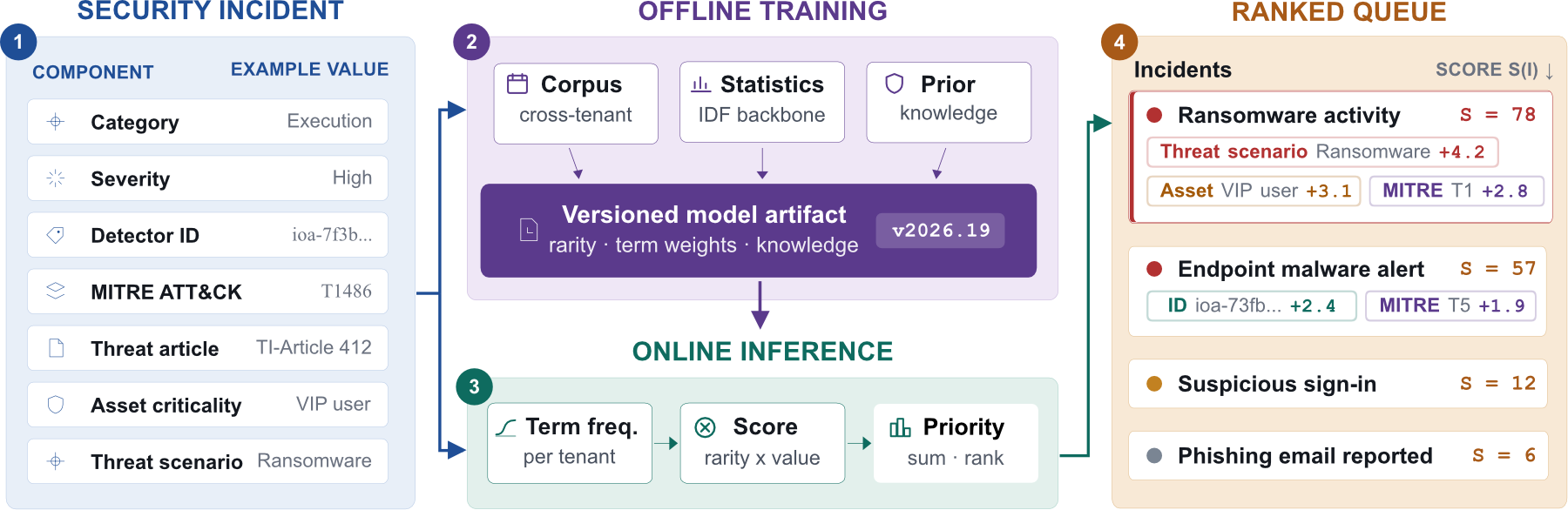}
\caption{Overview of the \methodshort{} architecture: an industry-scale framework for adaptive incident prioritization. A daily training pipeline processes historical incident telemetry to estimate global component rarity and produce a versioned model artifact. A near-real-time inference pipeline converts active incidents into security components, computes priority scores, and emits explainable rankings to Microsoft Defender customers.}
\label{fig:crown}
\end{figure*}

\vspace{1mm}\noindent 
\textbf{Incident prioritization at scale.} Building an industry-scale incident prioritization system presents five fundamental challenges:

\begin{enumerate}[topsep=4pt, leftmargin=*, itemsep=3pt]
    \item \textbf{Ranking beyond coarse severity.} Prioritization must distinguish incidents that share the same severity label but differ substantially in potential urgency and impact. 

    \item \textbf{Heterogeneous tenant data.} A production system must operate across tenants with different queue sizes, product coverage, detector vocabularies, custom detections, and metadata quality. 

    \item \textbf{Adaptable ranking.} Ranking should combine global security evidence with tenant-specific preferences, allowing customer feedback to upweight or downweight components based on what each SOC wants surfaced or suppressed.

    \item \textbf{Explainable by design.} Security analysts must understand why an incident moved up or down in the queue. Explanations should reflect the actual scoring logic and be inspectable by security researchers to build understanding and trust.

    \item \textbf{Low-latency.} Incident priority must refresh within active SOC triage timelines, continuously re-scoring queues as incidents arrive or change so analysts see updated priority within the 5-10 minute assessment windows expected by large SOCs.
\end{enumerate}

\subsection{Contributions}
We introduce \methodshort{} (Figure~\ref{fig:crown}), a retrieval-inspired framework designed to address the challenges of incident prioritization at scale. Our framework makes contributions in the following areas:

\begin{itemize}[topsep=2mm, itemsep=0mm, parsep=1mm, leftmargin=*]
    \item \textbf{\methodshort{} architecture.} \methodshort{} advances incident queue prioritization through a transparent ranking architecture designed in close collaboration with security research experts. Key innovations include: (1) representing incidents as bags of normalized security components; (2) adapting BM25-style ranking to a query-less SOC queue by combining local incident evidence with cross-tenant component rarity; and (3) incorporating bounded domain-prior multipliers that encode security-research knowledge while preserving component-level explanations. By disclosing these design and deployment details, we provide unique insight into an industry-scale incident prioritization system.
    
    \item \textbf{Extensive evaluation.} We evaluate \methodshort{} through expert-reviewed top-of-queue Precision@K, parameter sensitivity, score attribution, post-launch customer feedback and engagement telemetry, public benchmarking, and production runtime analysis. Across 1k customer organizations, \methodshort{} achieves 92.8\% P@10. In post-launch telemetry across 473K organization-day queues, \methodshort{} increases alert-detail interaction by 5.8\% and alert-detail view events by 17.5\% relative to severity ordering, providing behavioral evidence that model-ranked queues concentrate analyst engagement. 
    These results show that a lightweight, explainable, retrieval-inspired model can materially improve incident queue ordering while remaining practical for production workflows.
    
    \item \textbf{Public prioritization benchmark.} We extend the Microsoft \dataset{} dataset with the first public label source for SOC queue prioritization over real-world incidents. Existing \dataset{} labels support triage-style evaluation by identifying incident outcomes, but do not capture the operational queue-ordering question of which incidents analysts should investigate first. Our extension adds expert-derived priority labels for 9,980 incidents across 499 organization queues, establishing a foundation for future incident prioritization research.

    \item \textbf{Impact to Microsoft customers and beyond.} \methodshort{} is integrated into Microsoft Defender as the ranking algorithm behind Queue Assistant~\cite{gharib2026introducing,microsoft2026queueassistant}, and is deployed across tens of thousands of customers. 
    \methodshort{} advances Defender incident queues from recency- or severity-based ordering toward adaptive, evidence-driven prioritization, helping SOC analysts prioritize the most critical incidents for investigation. 
\end{itemize}

\section{Background}
We review four related areas of work: (1) incident correlation, (2) guided response and agentic SOC workflows, (3) alert and incident prioritization, and (4) retrieval-based ranking. 
Each area clarifies \methodshort{}'s scope: it does not correlate alerts into incidents, triage incidents as true or false positives, recommend remediation actions, or investigate incidents. Instead, \methodshort{} continuously ranks active incident queues.

\subsection{Incident Correlation}
Incident correlation systems connect related alerts into cohesive incident narratives~\cite{freitas2024graphweaver,elshoush2013intrusion,granadillo2016new,kotenko2022systematic}.
Correlation provides the substrate on which \methodshort{} operates, but addresses a different problem. Correlation determines which alerts belong together; prioritization determines how the resulting incidents should be ordered for analyst attention.

\subsection{Guided Response and Agentic Workflows}
Guided response and agentic SOC systems assist analysts after an incident has been selected for investigation. These systems provide similar-incident retrieval, triage support, query recommendations, evidence synthesis, and remediation guidance~\cite{freitas2025ai,jiang2024xpert,freitas2026genai,eilertsen2025towards,wei2025cortex,google2026use,banstola2026socai,lazer2026survey}. 
These systems are complementary to \methodshort{} but address a later stage of the SOC workflow. They help analysts determine what happened and what to do next once an incident receives attention; \methodshort{} addresses the preceding queue-management problem of deciding which incidents should receive attention first.

\subsection{Alert and Incident Prioritization}
Alert prioritization has long been studied as a way to reduce SOC workload and focus analysts on high-value security activity. Early prioritization approaches include rule-based ~\cite{alsubhi2008alert}, fuzzy-logic~\cite{alsubhi2012fuzmet}, and risk- or context-aware scoring~\cite{wang2024alertpro,jalalvand2025alert}. More recent systems apply machine learning to predict alert or incident actionability~\cite{gelman2023teq}, suppress false positives~\cite{oliver2024carbon,turcotte2025aact}, or prioritize investigation effort over high-value alerts~\cite{liu2022rapid}. Industry products also expose risk or priority scores, including risk-based alerting over entities~\cite{splunk2026risk}, and incident-scoring mechanisms based on manual, rule-based, and automated scores~\cite{paloalto2026incident,paloalto2023smartscore,crowdstrike2019crowdscore}.
Public alert-prioritization datasets typically operate at the alert, network-flow, or controlled attack level and rely on severity, triage, actionability, or synthetic priority labels~\cite{landauer2024introducing,ndichu2026ai,turcotte2025aact}. In contrast, our \dataset{} extension provides expert priority labels over real-world incident queues.

\subsection{Retrieval-Based Ranking}
Information retrieval models rank documents by combining local term evidence with corpus-level informativeness. TF-IDF weights terms by document frequency and corpus rarity~\cite{salton1988term}, log-entropy weighting downweights terms broadly distributed across documents~\cite{dumais1991improving}, and BM25 adds saturated term frequency to prevent repeated terms from dominating the score~\cite{robertson2009probabilistic}. These methods provide simple, additive ranking functions that are easy to inspect and operationalize.
\methodshort{} adapts this retrieval family to a query-less SOC setting: incidents are scored by their own security components rather than by relevance to a user query.
\section{Incident Ranking Methodology}\label{sec}
This section formalizes the scoring model used by \methodshort{} to rank active incidents by investigation priority. Unlike standard BM25, SOC prioritization has no explicit user query, so \methodshort{} adapts BM25 into a query-less incident score. Each incident is represented as a bag of normalized security components extracted from alert metadata.
The model reflects how analysts reason about priority: an incident is more likely to warrant attention when it contains multiple independent signals, rare or high-information evidence, and context suggesting meaningful impact. \methodshort{} captures these factors by combining saturated local component frequency, global component rarity, and bounded security-research priors. This allows informative evidence to accumulate across an incident while limiting the effect of repeated, common, or noisy signals.
The evaluated product experience uses the global domain-adjusted model. An optional tenant-preference overlay for future personalization is described in Section~\ref{subsec:tenant_adaptation}.

\subsection{Problem Formulation}
Let $\mathcal{T}$ be the set of tenants and let $\mathcal{I}=\bigcup_{t\in\mathcal{T}}\mathcal{I}_t$ be the global training corpus of incidents observed in the rolling training window. For a tenant $t$, let $Q_t$ denote its active incident queue. Each incident $i$ is represented as a multiset $C_i$ of normalized security components aggregated from its constituent alerts and metadata. Components are drawn from three prioritization-oriented families:

\begin{itemize}[topsep=2mm, itemsep=0mm, parsep=1mm, leftmargin=*]
    \item \textbf{Detection context:} product, severity, detector identifier, and detector quality tier.
    
    \item \textbf{Threat context:} MITRE ATT\&CK technique, threat-intelligence signal~\cite{freitas2025web}, and threat scenario tag.
    
    \item \textbf{Impact context:} attack-disruption, and critical assets such as domain controllers or VIP users. 
\end{itemize}

Heterogeneous incident metadata is normalized into categorical component keys so that signals from different products and alert schemas can be scored in a common representation. For example, severity values, detector identifiers, MITRE ATT\&CK techniques, threat-intelligence references, and critical-asset tags are mapped to component keys within their corresponding families (e.g., \texttt{severity:high}). The resulting multiset $C_i$ preserves repeated local evidence, while the set of unique components $U_i$ defines the scoring terms used by the ranking model.

Let $U_i$ be the set of unique components in $C_i$, and let $f_{c,i}$ be the frequency of component $c \in U_i$ in the multiset $C_i$. The prioritization task is to compute a non-negative ranking score $S_t(i)$ for every $i \in Q_t$ and rank incidents by descending score. The score depends on the incident's own components and the global model artifact, not on other incidents in $Q_t$. For analyst-facing surfaces, \methodshort{} also exposes a bounded display score $D_t(i)$ capped at 100, described in Section~\ref{subsec:score_explanation}. This display cap provides a stable product scale for analysts but does not affect queue ordering.

\subsection{Component Frequency}
We quantify the \emph{local} evidence for each component within an incident using a BM25-style saturated frequency. The goal is to reward components that appear in the incident while ensuring repeated occurrences of the same component provide diminishing returns, while also controlling how much incident breadth contributes to the score.
For each component $c$ in incident $i$, we compute:

\begin{equation}
\mathrm{tf}(c,i)=
\frac{f_{c,i}(k_1+1)}
{f_{c,i}+k_1\left(1-b+b\cdot\frac{\ell_i}{\bar{\ell}}\right)}
\label{eq:tf}
\end{equation}

where $f_{c,i}$ is the frequency of component $c$ in incident $i$, $\ell_i=\sum_{c'} f_{c',i}$ is the incident component length, and $\bar{\ell}$ is the average component length across the training corpus. The parameter $k_1$ controls how quickly repeated occurrences saturate, while $b$ controls the strength of incident-length normalization.
This normalization is useful because incident breadth has mixed semantics in SOC queues. Broad incidents may indicate ransomware, lateral movement, multi-stage attack progression, or impact across sensitive assets, but they may also reflect noisy detectors generating repetitive alerts. Rather than fully penalizing or fully rewarding breadth, \methodshort{} learns a middle ground through the $k_1$ and $b$ configuration: diverse informative components can still accumulate score, while repeated or low-value evidence is limited through frequency saturation, partial length normalization, domain-prior adjustment, and inference-time component capping. We evaluate sensitivity to $k_1$ and $b$ in Section~\ref{subsec:parameter_sensitivity}.

\subsection{Component Rarity}\label{sec:dk-idf}
We use \textit{global} component rarity to measure how informative a component is across the historical incident corpus. Let $N=|\mathcal{I}|$ be the number of training incidents and let $n_c$ be the number of training incidents containing component $c$. We compute smoothed inverse document frequency as:

\begin{equation}
\mathrm{idf}(c)=\log\left(\frac{N+1}{n_c+1}\right)
\label{eq:idf}
\end{equation}

The smoothed form is conservative for extremely rare components, which is important because singleton or near-singleton components can arise from experimental, custom, or third-party detections. 
Components unseen during inference are ignored for scoring until they are incorporated into a future training round, preventing previously unobserved metadata from arbitrarily increasing incident priority.

\medskip\noindent
\textbf{Domain-prior adjustment.}
Empirical rarity is useful for identifying informative components, but rarity alone does not imply urgency. This is especially true for custom and third-party detections, whose rules are authored outside our direct control and may attach rare metadata---such as MITRE ATT\&CK techniques---to broad, noisy, informational, or non-actionable alerts. In practice, domain knowledge is therefore used primarily to downweight low-value components, rather than to boost rare components.
We incorporate this security-research knowledge using a bounded multiplier over IDF. Let $\lambda_d(c)$ denote the domain-prior multiplier for component $c$. The domain-prior-adjusted rarity is:

\begin{equation}
\mathrm{idf}_{d}(c)=\lambda_d(c)\cdot\mathrm{idf}(c)
\label{eq:idfd}
\end{equation}

where $\lambda_d(c)=1$ by default, with values constrained to $[0.1,2]$.
This formulation preserves the interpretation of IDF as global rarity while allowing security researchers to adjust specific components. We disclose the formulation and bounds, but not the exact prior table because it encodes proprietary security-research knowledge and product-specific behavior.

\subsection{Tenant Preference Adaptation}\label{subsec:tenant_adaptation}
We enable tenant-specific adaptation through a bounded component-level preference overlay. This allows organizations to adjust prioritization toward the threat scenarios and workflows they care about most, while keeping each adjustment within fixed safety bounds. We model tenant adaptation as a bounded multiplier over domain-adjusted rarity. Let $\lambda_t(c)$ denote the tenant-preference multiplier for component $c$ in tenant $t$. By default, $\lambda_t(c)=1$, with allowed values constrained to $[0.5,2]$.
The tenant-adjusted rarity is:

\begin{equation}
\mathrm{idf}_{t}(c)=\lambda_t(c)\cdot\mathrm{idf}_{d}(c)
\label{eq:idft}
\end{equation}

In the feedback loop, analysts would provide component-level preferences, such as indicating that a contributing factor should be surfaced more or less often for their tenant. To make positive and negative feedback symmetric, we update the tenant multiplier in log space. Let $z_t(c)=\log \lambda_t(c)$. The update is:

\begin{equation}
z_t(c) \leftarrow
\operatorname{clip}{[\log(0.5),\log(2)]}
\left(z_t(c)+\eta y_{t,c}\right)
\label{eq:tenant_feedback_update}
\end{equation}

\begin{equation}
\lambda_t(c)=\exp(z_t(c))
\label{eq:tenant_multiplier}
\end{equation}

where $y_{t,c}\in\{-1,+1\}$ encodes negative or positive feedback for component $c$, and $\eta$ is the feedback step size. The update is symmetric because doubling and halving are equally distant in log space: $\log(2)=-\log(0.5)$. We set $\eta=\log(2)/6$, so six consecutive positive feedback events move a neutral multiplier from $1.0$ to the upper cap of $2.0$, while six consecutive negative events move it to the lower cap of $0.5$. 
This overlay is a tested backend capability but not enabled in the current experience; the deployed product and all main evaluations use the global domain-adjusted model.

\begin{figure}[b]
\centering
\includegraphics[width=\linewidth]{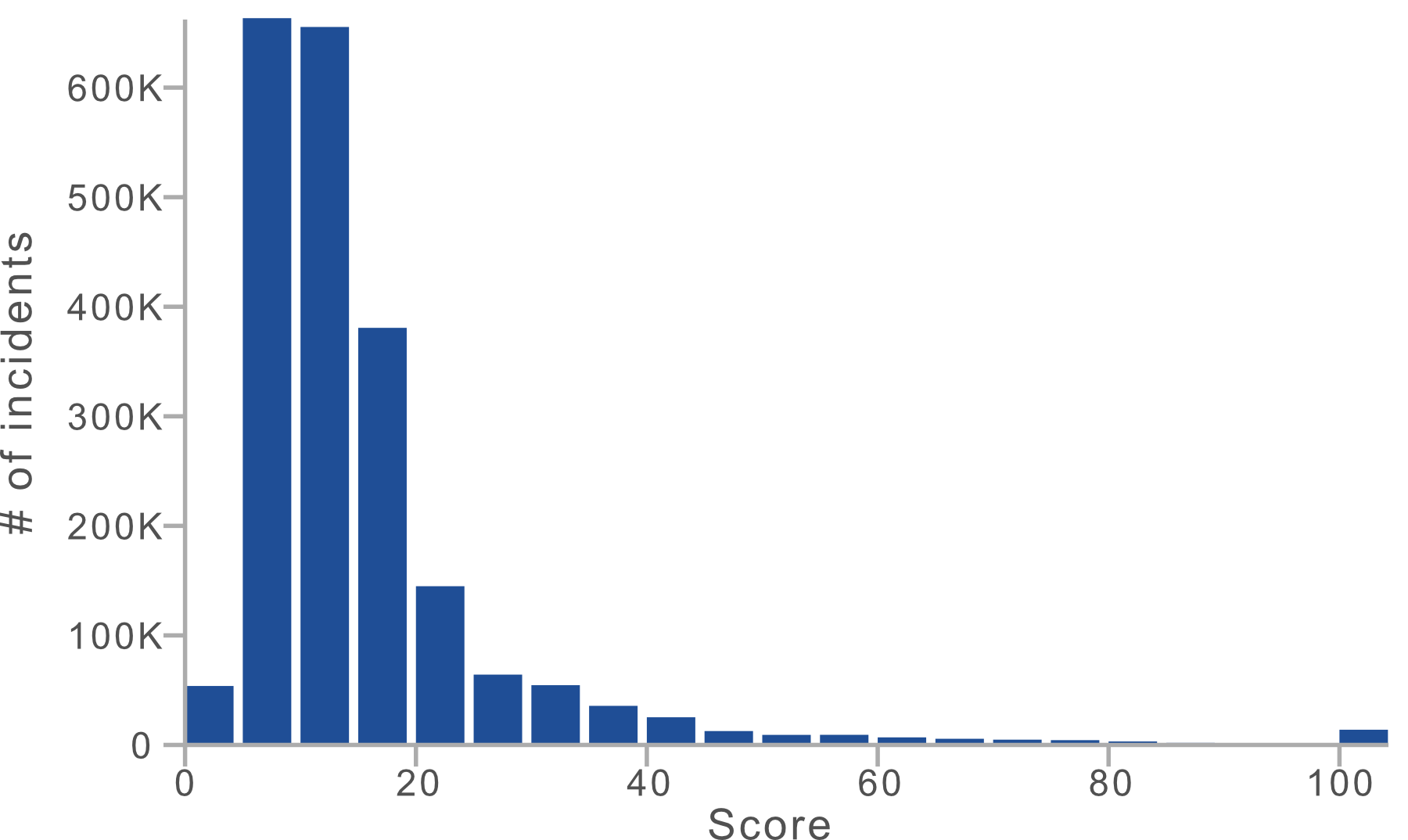}
\caption{Distribution of display scores across a one-week sample of production incidents. Scores show a long tail in which most incidents fall in low-to-middle ranges with a few approaching the display cap.}
\label{fig:score_distribution}
\vspace{-5mm}
\end{figure}

\subsection{Incident Score and Explanation}\label{subsec:score_explanation}
We now combine local component frequency, domain-adjusted rarity, and tenant preferences into a single incident score. The ranking score is additive over unique components:

\begin{equation}
S_t(i)=\sum_{c\in U_i}\mathrm{idf}_{t}(c)\cdot\mathrm{tf}(c,i)
\label{eq:score}
\end{equation}

When tenant adaptation is disabled or no tenant feedback is available, $\lambda_t(c)=1$ for all components, so $\mathrm{idf}_{t}(c)=\mathrm{idf}_{d}(c)$ and the score reduces to the global \methodshort{} score. The score assigned to each component is exactly:

\begin{equation}
s_t(c,i)=\mathrm{idf}_{t}(c)\cdot\mathrm{tf}(c,i)
\label{eq:component_score}
\end{equation}

Because the score is additive, explanations are faithful to the ranking function: the top-$k$ component scores $s_t(c,i)$ are exactly the largest scoring terms contributing to $S_t(i)$ and are shown to downstream product surfaces as priority factors.
Incidents are ranked by $S_t(i)$, preserving the full ordering resolution of the model, including distinctions among very high-priority incidents. BM25-style scores are unbounded and follow a long-tailed distribution: most incidents receive moderate scores, while a small number of unusually broad or high-signal incidents exceed 100, as shown in Figure~\ref{fig:score_distribution}. This long-tail behavior is useful for ranking, but less suitable for analyst-facing surfaces where scores should remain easy to interpret.
We therefore expose a bounded display score:

\begin{equation}
D_t(i)=\min\left(\mathrm{round}(S_t(i)),100\right)
\label{eq:display_score}
\end{equation}

where the 100-point cap provides a stable display scale, and aligns with analyst expectations that priority scores should fall within a fixed, interpretable interval. The cap is presentation-only and does not affect queue ordering, which uses the uncapped raw score $S_t(i)$. In a production sample across 108k organization queues, only 3.28\% had at least one capped incident. Among affected queues, cap concentration was modest: median 2, P75 4, and P90 8 capped incidents per queue.

\section{Production Deployment}\label{sec:deployment}
\methodshort{} is deployed globally as the ranking system behind Queue Assistant~\cite{gharib2026introducing,microsoft2026queueassistant}, serving tens of thousands of customer organizations through \platform{}. To comply with privacy and data-residency requirements, \methodshort{} is replicated across geographic regions and processes incident telemetry within the appropriate regional boundary.

\medskip
\noindent\textbf{Tenant isolation.}
\methodshort{} preserves tenant isolation throughout scoring and explanation generation. Each tenant's incidents are scored only from its own incident components and the regional model. Explanations expose only the components contributing to the scored incident for that tenant; exact customer data, tenant-specific feedback, and incident details are never exposed to other tenants.

\medskip
\noindent\textbf{Safety and evasion.}
\methodshort{} ranks incidents already emitted by the security platform and does not determine whether an attack is detected. However, prioritization can become an attack surface: an adversary could try to avoid high-weight signals, and cross-tenant rarity estimation could be distorted if a component is artificially flooded into the rolling training corpus. \methodshort{} mitigates these risks by relying on multiple component families and re-scoring incidents as new evidence arrives. Production monitoring tracks component-frequency shifts, unseen-component rates, and top-contributor changes, enabling suspicious activity to be handled through model rollback. 

\subsection{Training Pipeline}\label{sec:training}
The training pipeline runs daily over a rolling 30-day cross-tenant incident corpus. It converts each historical incident into normalized security components, computes the global document frequency $n_c$ for each component, estimates the corpus size $N$, and produces the IDF table in Equation~\ref{eq:idf}. The pipeline also applies the versioned domain-prior multiplier table, which allows security researchers to upweight high-risk components and downweight known low-value, test, sample, or noisy components.
Artifacts are versioned for reproducible rollout and rollback, and are consumed by the regional inference service after validation.

\medskip
\noindent\textbf{Infrastructure.} Training is implemented as a regionally replicated Azure Synapse pipeline over telemetry stored in ADLS. Synapse orchestrates deployment, monitoring, autoscaling, and retry behavior, while elastic PySpark pools perform large-scale component extraction and corpus-level aggregation within each geographic boundary. This design keeps training scalable across regions while providing operational safeguards against transient job failures.

\subsection{Inference Pipeline}\label{sec:inference}
For each scored incident, the pipeline applies the same component-extraction logic used during training and looks up each component's IDF and domain-prior multiplier. Before scoring, it applies an inference-time per-alert cap that retains only the three rarest components per component family, limiting the influence of over-specified custom or third-party alerts. We set the cap to three based on production alert statistics: across more than 23M alerts, the 95th percentile alert contains at most three MITRE ATT\&CK tags, while extreme outliers contain as many as 94. Thus, the cap preserves typical multi-tag alerts while bounding pathological cases where a single alert attaches many rare components.
The pipeline then applies any tenant-preference multiplier, computes saturated component frequencies, and sums component scores to produce the ranking score $S_t(i)$. This cap is inference-only and does not affect the training corpus used to estimate global component rarity.

\medskip
\noindent\textbf{Infrastructure.} Inference is performed per incident by a long-running containerized Python service. The service initializes scoring workers and loads the latest ranking artifacts at startup. For each configured inference job, workers consume incident-update messages from Azure Storage Queues, execute the corresponding Python inference job, and publish ranking and explanation results to downstream product storage. Incidents are re-scored whenever they are newly observed or updated, including when a new alert is added.

\section{Experiments}\label{sec:experiments}
We evaluate \methodshort{} using the protocol in Sec.~\ref{sec:eval_setup} and report results along six dimensions: BM25 parameter sensitivity for saturation and length normalization (Sec.~\ref{subsec:parameter_sensitivity}), expert-reviewed online queue ordering against operational and retrieval-style baselines (Sec.~\ref{sec:topk_precision}), score attribution across component families (Sec.~\ref{subsec:score_attribution}), the public \dataset{} prioritization benchmark (Sec.~\ref{subsec:guide_extension}), post-launch customer feedback (Sec.~\ref{subsec:customer_feedback}), and production runtime (Sec.~\ref{subsec:runtime}).

\medskip
\noindent\textbf{Scope and limitations.}
\methodshort{} prioritizes the existing incident queue; it does not replace triage systems that determine whether an incident is true or false positive, and it cannot surface attacks that were never emitted as incidents. Ranking quality depends on the availability and consistency of alert metadata, including MITRE ATT\&CK mappings, threat-intelligence references, and asset labels. The tenant-preference overlay provides a backend mechanism for future customer-specific adaptation, but it is not enabled in the current Defender Queue Assistant product experience. Our evaluation therefore focuses on the out-of-the-box global ranking experience that all customers receive by default.

\subsection{Evaluation Setup}\label{sec:eval_setup}
We evaluate production queues across regions with at least 50 incidents and 10 unique detector IDs. This ensures that evaluation queues are large and diverse enough to measure meaningful prioritization, rather than trivially small or homogeneous queues dominated by a select few detection patterns.

\medskip
\noindent\textbf{Review protocol.}
For each organization, expert security researchers create an independent reference ordering by performing a full-queue sweep over all active incidents in the organization queue and identifying which incidents should receive analyst attention first. The expert top-$20$ is therefore selected from the full queue, not from a union of method outputs, or a pre-filtered candidate pool.
Reviewers are blinded to \methodshort{} scores, baseline rankings, and method identities, and apply a fixed priority rubric that considers urgency, potential business impact, evidence of active or escalating compromise, affected scope, sensitive assets, containment status, and time-sensitive response needs. This produces ground-truth priority labels and rankings independent of \methodshort{} and all baselines. Evaluation is then performed automatically by comparing each method's ranked output against the expert-created ordering.

\medskip
\noindent\textbf{Baselines.}
We compare against operational baselines, including time ordering, highest detection-rule severity, alert-count ordering, and retrieval-style baselines such as TF-IDF, log-entropy, and vanilla BM25. We do not use learning-based actionability systems as direct baselines because they optimize different supervision targets, such as triage outcomes, rather than queue-relative urgency. Expert rankings are reserved for held-out evaluation rather than training because they are expensive to obtain across regions, tenants, detectors, and evolving queues. Our evaluation therefore compares methods that operate under the same unsupervised, cross-tenant production constraints as \methodshort{}.

\medskip
\noindent\textbf{Metrics.}
We report Precision@K because it directly matches the SOC queue objective: surfacing incidents that experts would prioritize for analyst attention. We evaluate $K\in\{5,10,20\}$ to capture the region where analyst attention is most concentrated. This product experience is set-oriented: analysts need the right incidents near the top of the queue, while the exact ordering within that visible set is less important. Since each reviewed organization contributes one queue, results are macro-averaged across queues, with 95\% confidence intervals computed using 10k paired bootstrap resamples.

\subsection{Parameter Sensitivity}\label{subsec:parameter_sensitivity}
We evaluate sensitivity to the BM25 saturation parameter $k_1$ and length-normalization parameter $b$ using a separate validation set of 261 queues. The sweep is performed under the deployed \methodshort{} scoring stack, with the inference-time component cap and domain-prior multipliers fixed to production settings; separate ablations in Table~\ref{tab:precision_baselines} isolate the effects of these controls. We sweep $k_1\in\{0,0.5,1.0,1.5,2.0\}$ and $b\in\{0,0.25,0.5,0.75,1.0\}$, where lower $k_1$ values make repeated evidence saturate more quickly and higher $b$ values apply stronger length normalization.
The complete parameter sweep is reported in Appendix~\ref{subsec:full_parameter_sweep}.

Table~\ref{tab:parameter_sensitivity} compares the selected configuration with representative ablations for binary term frequency, no length normalization, full length normalization, and stronger saturation. We select $k_1=2.0,b=0.5$ based on validation-set P@10 under the full \methodshort{} scoring stack. This configuration achieves the highest validation P@10 of 91.3\% (95\% CI: 88.7\%--93.7\%), with 88.3\% P@5 and 95.8\% P@20. The result suggests that repeated-but-saturated evidence and incident breadth are useful prioritization signals when combined with the deployed cap and domain-prior controls. Because detector behavior, alert schemas, and incident-correlation practices vary across security ecosystems, the ideal $k_1$ and $b$ settings may differ in other deployments. The selected configuration is fixed before evaluation on the held-out test queues in Section~\ref{sec:topk_precision}.

\begin{table}[t]
\centering
\renewcommand{\arraystretch}{1.05}
\begin{tabular*}{\columnwidth}{@{\extracolsep{\fill}}lccccc@{}}
\toprule
\textbf{Setting} & \textbf{$k_1$} & \textbf{$b$} & \textbf{P@5} & \textbf{P@10} & \textbf{P@20} \\
\midrule
\textbf{Best validation} & 2.0 & 0.50 & 88.3 & 91.3 & 95.8 \\
Binary TF & 0.0 & -- & 67.7 & 71.0 & 74.5 \\
No length norm. & 2.0 & 0.00 & 76.9 & 82.8 & 88.3 \\
Full length norm. & 2.0 & 1.00 & 70.6 & 76.5 & 82.6 \\
High saturation & 0.5 & 0.50 & 71.6 & 73.8 & 77.8 \\
Moderate saturation & 1.0 & 0.50 & 74.2 & 78.3 & 81.1 \\
\bottomrule
\end{tabular*}
\caption{Representative sensitivity of $k_1$ and $b$ across 261 validation queues. Inference-time component cap and domain-prior multipliers are held fixed to the deployed configuration; separate ablations in Table~\ref{tab:precision_baselines} isolate their contributions.}
\label{tab:parameter_sensitivity}
\vspace{-5mm}
\end{table}
\begin{table}[t]
\centering
\renewcommand{\arraystretch}{1.08}
\begin{tabular*}{\linewidth}{@{\extracolsep{\fill}}llccc}
\toprule
\textbf{Setting} & \textbf{Ordering} & \textbf{P@5} & \textbf{P@10} & \textbf{P@20} \\
\midrule
\multirow{10}{*}{Online}
& Random & 4.7 & 9.5 & 18.9 \\
& Time & 9.9 & 13.7 & 21.2 \\
& Max severity & 53.1 & 52.1 & 54.1 \\
& Alert count & 37.1 & 46.7 & 59.4 \\
& TF-IDF & 70.9 & 73.7 & 77.2 \\
& Log-entropy & 70.2 & 72.5 & 76.5 \\
& BM25 & 77.5 & 78.6 & 80.3 \\
& BM25 + cap & 78.1 & 79.3 & 81.0 \\
& BM25 + priors & 87.1 & 90.3 & 94.4 \\
& \methodshort{} & \textbf{89.2} & \textbf{92.8} & \textbf{96.9} \\
\midrule
\multirow{8}{*}{\dataset{}}
& Random & 5.4 & 10.9 & 21.7 \\
& Time & 4.2 & 7.7 & 14.9 \\
& Alert count & 20.3 & 28.1 & 38.9 \\
& TF-IDF & 64.6 & 70.0 & 74.8 \\
& Log-entropy & 55.8 & 60.1 & 64.2 \\
& BM25 & 93.4 & 96.3 & 98.7 \\
& BM25 + cap & \textbf{97.8} & \textbf{98.9} & \textbf{99.8} \\
\bottomrule
\end{tabular*}
\caption{Top-of-queue precision on online production queues and the public \dataset{} prioritization benchmark. Results are macro-averaged across reviewed organization queues. Random is the expected Precision@K from uniformly selecting $K$ incidents from the same full queue. Online ablations isolate the inference-time cap and domain-prior multipliers; \dataset{} reports only publicly reproducible variants.}
\label{tab:precision_baselines}
\vspace{-5mm}
\end{table}

\subsection{Online Queue Ordering}\label{sec:topk_precision}
We evaluate whether \methodshort{} improves the quality of incidents surfaced at the top of online production queues. Expert security researchers reviewed incident queues from 1000 customer organizations, stratified across geographic region, industry, and queue size.
Table~\ref{tab:precision_baselines} reports top-of-queue precision for \methodshort{} against operational and retrieval-style baselines. In online production queues, \methodshort{} achieves 92.8\% P@10 (95\% CI: 91.6\%--94.0\%). Operational baselines expose common queue-ordering limitations: time ordering can bury urgent incidents behind recent routine activity; severity ordering provides only coarse priority tiers that are inconsistently calibrated across custom and third-party detections; and alert-count ordering can over-rank noisy but low-value incidents. Retrieval-style baselines such as TF-IDF and log-entropy capture global component informativeness, while BM25 adds local saturation so repeated components contribute diminishing evidence. \methodshort{} performs best because it combines saturation, security-research priors, and inference-time component capping, allowing rare evidence to contribute while limiting known low-value, noisy, or over-specified components. Results are macro-averaged across reviewed queues, and uncertainty is computed using paired bootstrap resampling over queues.

\subsection{Score Attribution}\label{subsec:score_attribution}
\begin{figure}[t]
    \centering
    \includegraphics[width=\linewidth]{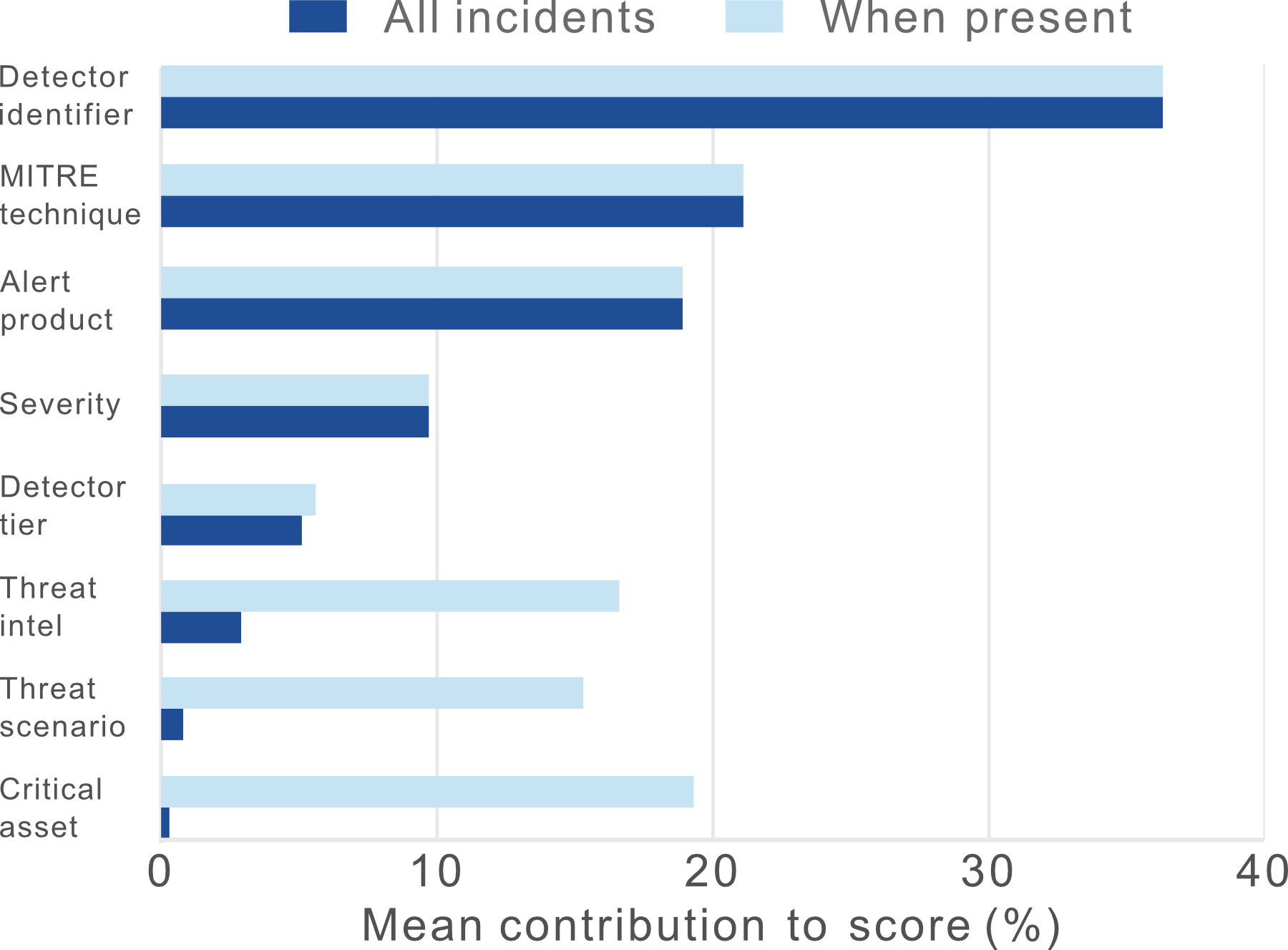}
    \caption{Contribution of component families to the \methodshort{} score across a production sample of 2.64M incidents from 99K organizations. 
    Contributions are averaged independently by family and therefore not constrained to sum to 100\%.
    }
    \label{fig:component_contribution}
\end{figure}

We analyze which component families contribute most to \methodshort{} scores in production. Figure~\ref{fig:component_contribution} reports mean score contribution across a sample of 2.64M production incidents from 99K organizations. We show two views: contribution averaged over all incidents, and contribution conditioned on the component family being present. Contributions are averaged independently by family and therefore are not constrained to sum to 100\%.
Detector identifiers contribute most broadly, reflecting that detection provenance is a strong prioritization signal. MITRE ATT\&CK techniques, alert product, and severity also provide substantial aggregate signal because they appear across many incidents. In contrast, threat-intelligence, threat-scenario, and critical-asset components appear less frequently, but contribute strongly when present. This matches the intended behavior of the model: common detection context provides broad ranking signal, while rarer high-value security context can meaningfully raise priority when it appears.

\subsection{GUIDE Prioritization Benchmark}\label{subsec:guide_extension}
The public Microsoft \dataset{} dataset contains more than 13 million data points across 33 entity types, including 1.6 million alerts and 1 million incidents across 6.1k organizations~\cite{freitas2025ai}. While \dataset{} provides a large-scale real-world benchmark for SOC triage and remediation-action recommendation under the permissive CDLA-2.0 license, it does not support evaluation of incident prioritization. We address this gap by releasing an incident-prioritization label file over \dataset{} to support reproducible queue-ordering experiments.

\medskip\noindent
\textbf{Prioritization labels.}
We release the prioritization labels on \href{https://www.kaggle.com/datasets/Microsoft/microsoft-security-incident-prediction}{Kaggle}. They capture queue-relative priority rather than triage outcome, ranking which incidents within each organization should receive analyst attention first. The label file contains \texttt{OrgId}, \texttt{IncidentId}, \texttt{Rank}, and \texttt{Split}; identifiers link to the scrubbed \dataset{} records, while \texttt{Rank} provides the expert-derived within-queue ordering. Although we use Precision@K as the primary metric, the full rankings support rank-aware evaluation. Splits are assigned at the organization level to prevent cross-split leakage.

To construct these labels, we use incident context available during the creation of \dataset{}, allowing expert reviewers to assess priority with complete security context while releasing only scrubbed identifiers and priority rankings. We include organizations from the \dataset{} test split with at least 50 incidents and 10 distinct detector identifiers, yielding 499 organization queues. For each organization, experts perform a full-queue sweep over all eligible incidents and identify the top-20 incidents that should receive analyst attention first, producing 9,980 labeled incidents in total. Because queues are reviewed independently of AIP and all baselines, the released labels do not depend on any evaluated ranking method. Experts review incidents according to the rubric in Section~\ref{sec:eval_setup}, with disagreements adjudicated to produce final rankings.

\medskip\noindent
\textbf{Ranking results.}
Table~\ref{tab:precision_baselines} reports results on the public \dataset{} prioritization benchmark. Each method ranks candidate incidents within each organization queue. We compute Precision@K by comparing the method's top-$K$ incidents with the expert-ranked top-$K$ incidents for the same queue, then macro-average across organization queues. Because the proprietary domain-prior table is not released with \dataset{}, the benchmark reports the publicly reproducible \methodshort{} variant, BM25 + cap, rather than the full deployed production model. This variant achieves 98.9 P@10 with a 95\% CI: [98.4, 99.4].

The higher absolute P@K values on \dataset{} are not directly comparable to online production precision. \dataset{} is a fixed historical snapshot from an earlier product and detector ecosystem, before today’s broader custom and third-party integrations made live queues more heterogeneous and noisy. It also supports only a reduced public-metadata variant using fields such as detector identifiers and MITRE ATT\&CK techniques, whereas production evaluation reflects current tenant mix, detector drift, evolving product coverage, and richer proprietary signals. We therefore use \dataset{} as a reproducible benchmark for comparing methods within the released setting, not as a reproduction of the production system.
We encourage the community to explore additional metadata available in \dataset{} and develop alternative ranking methods under the released protocol.

\subsection{Online Customer Feedback}\label{subsec:customer_feedback}
We complement expert queue review with three post-launch operational signals: customer survey feedback, queue-ordering retention, and analyst engagement telemetry.

\medskip\noindent
\textbf{Customer survey feedback.}
Table~\ref{tab:customer_feedback} summarizes post-launch feedback from 17 customers with diverse queue configurations and sizes. Respondents reported that the feature worked as expected, with an average rating of 4.47 out of 5, and expressed positive overall satisfaction with Queue Assistant, with an average rating of 4.12 out of 5. Customers also reported that the scores reflected incident priority: 10 of 17 answered yes, 6 answered partially, and only 1 answered no. The strongest signal was explanation quality: all 17 respondents reported that the priority factors helped them understand the reasoning behind the score. This explanation-quality result reflects a core design goal of \methodshort{}: the displayed priority factors are exact component-level scoring terms, not post-hoc rationales. We treat this feedback as operational validation rather than unbiased ground truth because the sample is modest, respondents are self-selected, and responses reflect only customers exposed to the feature.

\medskip\noindent
\textbf{Queue-ordering retention.}
Because \methodshort{} is the default queue ordering, we distinguish default retention from explicit sort overrides. For each eligible queue session, we reconstruct the sort-state timeline from queue-open and manual-sort telemetry, excluding sessions with fewer than 10 incidents or with under 10 seconds of dwell time and no queue interaction. In the broad sample of organizations active on at least 10 of 30 days, the analysis covers 354{,}403 sessions from 4{,}266 organizations. Users retained the default \methodshort{} ordering in 97.4\% of sessions and explicitly sorted away from it in only 2.6\%, indicating that analysts rarely overrode the model-ranked queue. These trends are consistent under stricter 15- and 20-day activity thresholds. We interpret these metrics as behavioral adoption of the deployed ranking interface, not causal evidence of productivity or incident value.

\medskip\noindent
\textbf{Analyst engagement diagnostic.}
We evaluate whether \methodshort{} concentrates analyst attention on incidents that analysts subsequently inspect. For each organization and day, we construct an eligible incident queue from production-scored incidents, retaining incidents with either a model score $\geq 80$ or ``high'' severity. We then compare the \methodshort{}-ranked top-10 against a counterfactual severity-ranked top-10 over the same eligible queue. Engagement is measured using authenticated alert-detail views mapped back to incidents, indicating that an analyst inspected incident evidence beyond the queue listing. The analysis covers a production sample of 285 region-day buckets, 92,094 region-scoped organizations, 473K organization-day queues, and 10.9M incident-day candidates. Relative to severity ordering, \methodshort{} yields a 5.8\% relative lift in incident-level alert-detail interaction, with a 10K organization-day clustered bootstrap 95\% CI of 5.4\%--6.1\%. It also produces 17.5\% more alert-detail view events than severity ordering, with 95\% CI 15.8\%--19.5\%, suggesting both broader and deeper engagement with surfaced incidents.

The \methodshort{} and severity top-10 lists overlap on 79.6\% of placements. Among the 20.4\% of placements where the two rankings disagree, \methodshort{}-selected incidents have a 56.3\% higher alert-detail interaction rate than severity-selected incidents, with 95\% CI 52.0\%--60.7\%. These \methodshort{}-only placements also generate 262\% more alert-detail view events than severity-only placements, with 95\% CI 233\%--295\%, reflecting substantially greater view-event volume among incidents surfaced only by \methodshort{}. Because the telemetry is observational and collected under the deployed ranking---so engagement reflects both incident selection and display position---these results should be interpreted as behavioral evidence of analyst engagement, not as a causal estimate of analyst productivity or incident value.

\begin{table}[t]
\centering
\renewcommand{\arraystretch}{1.08}
\begin{tabular*}{\linewidth}{@{\extracolsep{\fill}}lccccc}
\toprule
\textbf{Question} & \textbf{Rating} & \textbf{Yes} & \textbf{Partial} & \textbf{Unsure} & \textbf{No} \\
\midrule
Worked as expected & 4.47 / 5 & -- & -- & -- & -- \\
Overall satisfaction & 4.12 / 5 & -- & -- & -- & -- \\
\midrule
Scores reflect priority & -- & 10 & 6 & -- & 1 \\
Factors explain score & -- & 17 & -- & -- & -- \\
\bottomrule
\end{tabular*}
\caption{Post-launch customer feedback for Defender Queue Assistant ($n=17$). Rating questions use a 1--5 scale; categorical questions report response counts.}
\label{tab:customer_feedback}
\vspace{-5mm}
\end{table}

\subsection{Production Runtime}\label{subsec:runtime}
We evaluate whether \methodshort{} satisfies the latency and scalability requirements of a production queue ranking service using telemetry across a representative sample of production regions. The daily training pipeline processes a rolling 30-day corpus, produces a median vocabulary of 8.7K unique components per region, and completes in 10.7 minutes at P50, 26.1 minutes at P90, and 32.1 minutes at P95, across regional runs. The resulting model artifact is versioned for reproducible rollback.
Inference runs at the incident level in real time: when an incident is newly observed or updated, including when a new alert is added, \methodshort{} extracts the incident components, loads the latest approved artifact, computes the raw and display scores, and emits updated ranking and explanation metadata. Per-incident scoring latency is 5 seconds at P50, 32 seconds at P90, and 92 seconds at P95.
The median-to-tail latency gap reflects variation in incident size and bursty updates; however, P95 scoring latency remains 92 seconds, comfortably below the 5-10 minute assessment windows expected by large SOCs.

\section{Conclusion}\label{sec:conclusion}
We introduced \method{} (\methodshort{}), an explainable, retrieval-inspired framework for prioritizing SOC incident queues at production scale. Deployed in Microsoft Defender, \methodshort{} adapts BM25-style ranking to a query-less security setting by representing incidents as bags of normalized security components.
By disclosing key architectural and deployment details, we provide unique insight into the design of an industry-scale incident prioritization system.
In expert-reviewed evaluation across 1,000 customer organizations, \methodshort{} achieves 92.8\% P@10. Post-launch telemetry across 473K organization-day queues shows a 5.8\% increase in alert-detail interaction and 17.5\% more alert-detail view events relative to severity ordering, providing behavioral evidence that \methodshort{} concentrates analyst engagement on model-ranked incidents. These results demonstrate that a lightweight and interpretable ranking model can substantially improve queue ordering while satisfying the latency, scalability, and explainability requirements of operational SOC workflows.
We also extend the Microsoft \dataset{} dataset with the first public label source for SOC queue prioritization over real-world incidents, covering 499 organization queues and 9,980 expert-ranked incidents. By releasing these labels, we establish a reproducible foundation for developing and comparing future incident-prioritization methods.

\begin{acks}
We thank all of our colleagues who supported this research.
\end{acks}

\bibliographystyle{ACM-Reference-Format}
\bibliography{main.bib}

@inproceedings{freitas2024graphweaver,
  title={GraphWeaver: Billion-Scale Cybersecurity Incident Correlation},
  author={Freitas, Scott and Gharib, Amir},
  booktitle={proceedings of the 33rd ACM international CIKM},
  year={2024}
}

@inproceedings{freitas2025ai,
  title={AI-driven guided response for security operation centers with Microsoft Copilot for Security},
  author={Freitas, Scott and Kalajdjieski, Jovan and Gharib, Amir and McCann, Robert},
  booktitle={Companion Proceedings of the ACM on Web Conference 2025},
  pages={191--200},
  year={2025}
}

@inproceedings{freitas2025web,
  title={Web scale graph mining for cyber threat intelligence},
  author={Freitas, Scott and Gharib, Amir},
  booktitle={Proceedings of the 31st ACM SIGKDD Conference on Knowledge Discovery and Data Mining V. 2},
  pages={4447--4456},
  year={2025}
}

@misc{google2026use,
  title={Use Triage and Investigation Agent to investigate alerts},
  author={Google},
  year={2026},
  institution={Google},
  url={https://docs.cloud.google.com/chronicle/docs/secops/triage-investigation-agent}
}

@misc{einav2023introducing,
    title={Introducing a Unified Security Operations Platform with Microsoft Sentinel and Defender XDR},
    author={Einav, Erez},
    year={2023},
    website={https://techcommunity.microsoft.com/blog/microsoftsentinelblog/introducing-a-unified-security-operations-platform-with-microsoft-sentinel-and-d/3983341}
}

@misc{gharib2026introducing,
  title={Introducing AI-powered incident prioritization in Microsoft Defender},
  author={Gharib, Amir and Freitas, Scott and Magenheim, Maayan},
  year={2026},
  institution={Microsoft},
  url={https://techcommunity.microsoft.com/blog/microsoftthreatprotectionblog/introducing-ai-powered-incident-prioritization-in-microsoft-defender/4483834}
}

@inproceedings{granadillo2016new,
  title={New types of alert correlation for security information and event management systems},
  author={Granadillo, Gustavo Gonzalez and El-Barbori, Mohammed and Debar, Herve},
  booktitle={2016 8th IFIP international conference on new technologies, mobility and security (NTMS)},
  pages={1--7},
  year={2016},
  organization={IEEE}
}

@inproceedings{elshoush2013intrusion,
  title={Intrusion alert correlation framework: An innovative approach},
  author={Elshoush, Huwaida Tagelsir and Osman, Izzeldin Mohamed},
  booktitle={IAENG Transactions on Engineering Technologies: Special Volume of the World Congress on Engineering 2012},
  pages={405--420},
  year={2013},
  organization={Springer}
}

@article{kotenko2022systematic,
  title={Systematic literature review of security event correlation methods},
  author={Kotenko, Igor and Gaifulina, Diana and Zelichenok, Igor},
  journal={IEEE Access},
  volume={10},
  pages={43387--43420},
  year={2022},
  publisher={IEEE}
}

@inproceedings{jiang2024xpert,
  title={Xpert: Empowering incident management with query recommendations via large language models},
  author={Jiang, Yuxuan and Zhang, Chaoyun and He, Shilin and Yang, Zhihao and Ma, Minghua and Qin, Si and Kang, Yu and Dang, Yingnong and Rajmohan, Saravan and Lin, Qingwei and others},
  booktitle={Proceedings of the IEEE/ACM 46th International Conference on Software Engineering},
  pages={1--13},
  year={2024}
}

@article{oliver2024carbon,
  title={Carbon Filter: Real-time Alert Triage Using Large Scale Clustering and Fast Search},
  author={Oliver, Jonathan and Batta, Raghav and Bates, Adam and Inam, Muhammad Adil and Mehta, Shelly and Xia, Shugao},
  journal={arXiv preprint arXiv:2405.04691},
  year={2024}
}

@article{lazer2026survey,
  title={A Survey of Agentic AI and Cybersecurity: Challenges, Opportunities and Use-case Prototypes},
  author={Lazer, Sahaya Jestus and Aryal, Kshitiz and Gupta, Maanak and Bertino, Elisa},
  journal={arXiv preprint arXiv:2601.05293},
  year={2026}
}

@inproceedings{eilertsen2025towards,
  title={Towards Agentic Investigation of Security Alerts},
  author={Eilertsen, Even and Mavroeidis, Vasileios and Grov, Gudmund},
  booktitle={2025 IEEE International Conference on Big Data (BigData)},
  pages={7793--7802},
  year={2025},
  organization={IEEE}
}

@article{wei2025cortex,
  title={CORTEX: Collaborative LLM Agents for High-Stakes Alert Triage},
  author={Wei, Bowen and Tay, Yuan Shen and Liu, Howard and Pan, Jinhao and Luo, Kun and Zhu, Ziwei and Jordan, Chris},
  journal={arXiv preprint arXiv:2510.00311},
  year={2025}
}

@inproceedings{banstola2026socai,
title={Experiences of Using Agentic AI to Fill Tooling Gaps in a Security Operations Center},
author={Banstola, Kritan and Al Faisal, Faayed and Ou, Xinming},
booktitle={Workshop on Security Operation Center Operations and Construction (WOSOC)},
year={2026}
}

@article{freitas2026genai,
  title={GenAI-Driven Threat Detection with Microsoft Security Copilot},
  author={Freitas, Scott and Gharib, Amir},
  journal={arXiv preprint arXiv:2605.20896},
  year={2026}
}

@inproceedings{alsubhi2008alert,
  title={Alert prioritization in Intrusion Detection Systems},
  author={Alsubhi, Khalid and Al-Shaer, Ehab and Boutaba, Raouf},
  booktitle={2008 IEEE Network Operations and Management Symposium},
  pages={33--40},
  year={2008},
  organization={IEEE}
}

@article{alsubhi2012fuzmet,
  title={FuzMet: A fuzzy-logic based alert prioritization engine for intrusion detection systems},
  author={Alsubhi, Khalid and Aib, Issam and Boutaba, Raouf},
  journal={International Journal of Network Management},
  volume={22},
  number={4},
  pages={263--284},
  year={2012},
  publisher={Wiley}
}

@article{jalalvand2025alert,
  title={Alert Prioritisation in Security Operations Centres: A Systematic Survey on Criteria and Methods},
  author={Jalalvand, Fatemeh and Baruwal Chhetri, Mohan and Nepal, Surya and Paris, C{\'e}cile},
  journal={ACM Computing Surveys},
  volume={57},
  number={2},
  pages={42:1--42:36},
  year={2025},
  publisher={ACM},
  doi={10.1145/3695462}
}

@article{gelman2023teq,
  title={That Escalated Quickly: An ML Framework for Alert Prioritization},
  author={Gelman, Ben and Taoufiq, Salma and V{\"o}r{\"o}s, Tam{\'a}s and Berlin, Konstantin},
  journal={arXiv preprint arXiv:2302.06648},
  year={2023}
}

@inproceedings{liu2022rapid,
  title={RAPID: Real-Time Alert Investigation with Context-aware Prioritization for Efficient Threat Discovery},
  author={Liu, Yushan and Shu, Xiaokui and Sun, Yixin and Jang, Jiyong and Mittal, Prateek},
  booktitle={Proceedings of the 38th Annual Computer Security Applications Conference},
  pages={827--840},
  year={2022}
}

@article{wang2024alertpro,
  title={Combating alert fatigue with AlertPro: Context-aware alert prioritization using reinforcement learning for multi-step attack detection},
  author={Wang, Xiaoyu and Yang, Xiaobo and Liang, Xueping and Zhang, Xiu and Zhang, Wei and Gong, Xiaorui},
  journal={Computers \& Security},
  volume={137},
  pages={103583},
  year={2024},
  publisher={Elsevier},
  doi={10.1016/j.cose.2023.103583}
}

@article{turcotte2025aact,
  title={Automated Alert Classification and Triage (AACT): An Intelligent System for the Prioritisation of Cybersecurity Alerts},
  author={Turcotte, Melissa and Labr{\`e}che, Fran{\c{c}}ois and Paquette, Serge-Olivier},
  journal={arXiv preprint arXiv:2505.09843},
  year={2025}
}

@misc{microsoft2026queueassistant,
  title={Prioritize incidents in the Microsoft Defender portal},
  author={Microsoft},
  year={2026},
  institution={Microsoft},
  url={https://learn.microsoft.com/en-us/defender-xdr/incident-queue}
}

@misc{paloalto2026incident,
  title={Learn about the different incident scoring methods},
  author={{Palo Alto Networks}},
  year={2026},
  institution={Palo Alto Networks},
  url={https://docs-cortex.paloaltonetworks.com/r/Cortex-XDR/Cortex-XDR-3.x-Documentation/Incident-scoring}
}

@misc{splunk2026risk,
  title={Risk scoring in Splunk Enterprise Security},
  author={Splunk},
  year={2026},
  institution={Splunk},
  url={https://help.splunk.com/en/splunk-enterprise-security-8/administer/8.5/risk-based-alerting}
}

@article{robertson2009probabilistic,
  title={The Probabilistic Relevance Framework: BM25 and Beyond},
  author={Robertson, Stephen and Zaragoza, Hugo},
  journal={Foundations and Trends in Information Retrieval},
  volume={3},
  number={4},
  pages={333--389},
  year={2009},
  publisher={Now Publishers},
  doi={10.1561/1500000019}
}

@article{salton1988term,
title={Term-weighting approaches in automatic text retrieval},
author={Salton, Gerard and Buckley, Christopher},
journal={Information Processing \& Management},
volume={24},
number={5},
pages={513--523},
year={1988},
publisher={Elsevier}
}

@inproceedings{dumais1991improving,
title={Improving the retrieval of information from external sources},
author={Dumais, Susan T},
booktitle={Behavior Research Methods, Instruments, \& Computers},
volume={23},
number={2},
pages={229--236},
year={1991}
}

@inproceedings{landauer2024introducing,
  title={Introducing a new alert data set for multi-step attack analysis},
  author={Landauer, Max and Skopik, Florian and Wurzenberger, Markus},
  booktitle={Proceedings of the 17th cyber security experimentation and test workshop},
  pages={41--53},
  year={2024}
}

@article{ndichu2026ai,
  title={AI-Driven Security Alert Screening and Alert Fatigue Mitigation in Security Operations Centers: A Comprehensive Survey},
  author={Ndichu, Samuel and Yamada, Akira and Ban, Tao and Ozawa, Seiichi and Takahashi, Takeshi and Inoue, Daisuke},
  journal={arXiv preprint arXiv:2605.08316},
  year={2026}
}

@misc{paloalto2023smartscore,
  title = {Unlocking the Black Box: Transparency for ML-Based Incident Risk Scoring},
  author = {{Palo Alto Networks}},
  year = {2023},
  howpublished = {\url{https://www.paloaltonetworks.com/blog/security-operations/unlocking-the-black-box-transparency-for-ml-based-incident-risk-scoring/}},
  note = {Accessed 2026-06-24}
}

@misc{crowdstrike2019crowdscore,
  title = {CrowdScore Dramatically Reduces Alert Fatigue},
  author = {{CrowdStrike}},
  year = {2019},
  howpublished = {\url{https://www.crowdstrike.com/en-us/blog/crowdscore-dramatically-reduces-alert-fatigue/}},
  note = {Accessed 2026-06-24}
}

\clearpage
\appendix
\renewcommand{\thesubsection}{\Alph{subsection}}

\noindent
\begin{minipage}{\columnwidth}

\section*{Appendix}
\label{sec:appendix}

\subsection{Full Parameter Sweep}
\label{subsec:full_parameter_sweep}

Table~\ref{tab:parameter_sweep_full} reports the complete $k_1$ and $b$ parameter sweep summarized in Section~\ref{subsec:parameter_sensitivity}.

\begin{table}[H] 
\centering
\renewcommand{\arraystretch}{1.05}
\begin{tabular*}{\columnwidth}{@{\extracolsep{\fill}}ccccc@{}}
\toprule
\textbf{$k_1$} & \textbf{$b$} & \textbf{P@5} & \textbf{P@10} & \textbf{P@20} \\
\midrule
0.0 & 0.00 & 67.7 & 71.0 & 74.5 \\
0.0 & 0.25 & 67.7 & 71.0 & 74.7 \\
0.0 & 0.50 & 67.7 & 71.1 & 74.7 \\
0.0 & 0.75 & 67.7 & 70.9 & 74.6 \\
0.0 & 1.00 & 67.2 & 70.5 & 74.3 \\
\midrule
0.5 & 0.00 & 71.9 & 74.5 & 78.9 \\
0.5 & 0.25 & 71.3 & 74.8 & 78.7 \\
0.5 & 0.50 & 71.6 & 73.8 & 77.8 \\
0.5 & 0.75 & 70.0 & 73.0 & 76.3 \\
0.5 & 1.00 & 66.8 & 70.4 & 74.0 \\
\midrule
1.0 & 0.00 & 73.3 & 76.7 & 80.6 \\
1.0 & 0.25 & 76.0 & 78.7 & 82.1 \\
1.0 & 0.50 & 74.2 & 78.3 & 81.1 \\
1.0 & 0.75 & 72.1 & 75.6 & 78.4 \\
1.0 & 1.00 & 66.1 & 70.7 & 73.7 \\
\midrule
1.5 & 0.00 & 71.7 & 75.1 & 78.9 \\
1.5 & 0.25 & 74.9 & 77.6 & 80.6 \\
1.5 & 0.50 & 75.8 & 78.4 & 80.6 \\
1.5 & 0.75 & 71.6 & 74.9 & 78.3 \\
1.5 & 1.00 & 63.8 & 68.8 & 72.5 \\
\midrule
2.0 & 0.00 & 76.9 & 82.8 & 88.3 \\
2.0 & 0.25 & 82.8 & 87.7 & 92.0 \\
\textbf{2.0} & \textbf{0.50} & \textbf{88.3} & \textbf{91.3} & \textbf{95.8} \\
2.0 & 0.75 & 81.1 & 85.3 & 89.8 \\
2.0 & 1.00 & 70.6 & 76.5 & 82.6 \\
\bottomrule
\end{tabular*}
\caption{Full $k_1$ and $b$ parameter sweep across 261 validation queues, under the deployed \methodshort{} scoring stack with the inference-time component cap and domain-prior multipliers held fixed. The selected configuration ($k_1=2.0$, $b=0.5$), used throughout the main paper, is shown in bold and achieves the highest validation P@10.}
\label{tab:parameter_sweep_full}
\vspace{-5mm}
\end{table}

\end{minipage}

\end{document}